\documentstyle[12pt,aps,epsf,eqsecnum]{revtex}

\begin{document}
\newcommand{\hgo}{$^{190}$Hg}
\newcommand{\hgt}{$^{192}$Hg}
\newcommand{\hgf}{$^{194}$Hg}
\newcommand{\J}{{\cal J}^{(2)}}
\newcommand{\w}{\omega_{\rm rot}}
\newcommand{\ep}{\epsilon}
\newcommand{\bra}[1]{\langle #1 |}
\newcommand{\ket}[1]{| #1 \rangle}

\draft

\title{Octupole Vibrations at High Angular Momenta\footnote{
Invited talks presented at
{\sl XXIV Mazurian Lakes School of Physics},
August, 1995, Piaski, Poland}}
\author{Takashi~Nakatsukasa\footnote{
E-mail : {\tt nakatsukasat@crl.aecl.ca}}}
\address{AECL, Chalk River Laboratories, Chalk River,
Ontario K0J 1J0, Canada}
\date{Preprint : TASCC-P-95-26}

\maketitle

\tighten
\bigskip
\bigskip
\begin{abstract}
Properties of octupole vibrations in rapidly rotating nuclei
are discussed.
Microscopic RPA calculations based on the cranked shell model are
performed to investigate
the interplay between rotation and vibrations.
The ability of this model
to describe the properties of
collective vibrations
built on the ground bands
in rare-earth and actinide nuclei is demonstrated
at high angular momentum.
The octupole vibrational states in even-even superdeformed Hg nuclei are
also predicted and compared with available experimental data.
A new interpretation
of the observed excited superdeformed bands
invoking these octupole bands
is proposed.
\end{abstract}

\pacs{PACS number(s): 21.10.Re, 21.60.-n, 23.20.-g}
\bigskip

\section{I\lowercase{ntroduction}}
\label{sec: intro}

The rapid rotation of a nucleus provides us
with an opportunity of studying
a quantum-many-body system with a strong Coriolis field.
There, various phenomena are expected to be observed;
the pairing phase transition (Mottelson-Valatin effect),
gapless superconductivity (back-bending),
shape transitions, yrast traps, etc.\cite{Ham85}.
Some of these are analogous to phenomena of
condensed-matter physics in
a strong magnetic field, while some are
peculiar to a {\it finite} system like the nucleus.
As an example of unique elementary excitations in the
{\rm finite} system,
surface-vibrational motion in a rapidly rotating nucleus is
of great interest.
In this talk the effects of the strong Coriolis field on octupole
vibrations are discussed in terms of a microscopic
model based on the cranked mean field extended by
the random-phase approximation (RPA).

Properties of unknown collective vibrations
built on the superdeformed (SD) yrast band can be also
predicted in this theoretical approach.
Since the large deformation and the rapid rotation of SD bands may
produce a novel shell structure,
quite different surface vibrational features are expected.
Experimental data suggesting vibrational-excited SD bands have been
recently reported for $^{190}$Hg\cite{Cro94}
and for $^{152}$Dy\cite{Dag95,Nak95}.
Octupole vibrations in SD Hg nuclei are discussed in the last section.

\section{M\lowercase{icroscopic} RPA \lowercase{in the rotating frame}}

In the cranked shell model extended by the RPA theory,
vibrational excitations built on the rotating vacuum are microscopically
described by
superpositions of a large number of two-quasiparticle states.
The RPA also allows us to describe non-collective
two-quasiparticle excitations
and weakly collective states which are difficult to discuss
by means of macroscopic models.
Effects of the Coriolis force on these various modes of excitation
are automatically taken into account in RPA solutions
since the mean field is affected by the cranking term
$-\hbar\omega_{\rm rot}J_x$.
Shortcomings of this model are the semi-classical treatment of rotation
(angular momentum vector) and the small-amplitude nature of the RPA method.
Thus we need to pay special attention when discussing properties of
low-spin states (see below).

The model Hamiltonian is assumed to be of the form:
\begin{equation}
 \label{hamiltonian}
 H'= h'_{\rm s.p.} + H_{\rm int}\ ,
\end{equation}
where $h'_{\rm s.p.}$ is a cranked single-particle Nilsson Hamiltonian.
The residual interactions are separable multipole interactions,
\begin{equation}
 \label{residual}
 H_{\rm int} = -\frac{1}{2}\sum_K \chi_{3K} Q_{3K}^\dagger Q_{3K}
       + \frac{1}{2}\sum_K \chi_{1K} \left(\tau_3 D_{1K}\right)^\dagger
                          \left(\tau_3 D_{1K}\right) \ ,
\end{equation}
where octupole (dipole) operators,
$Q_{3K}\equiv (r^{''})^3Y_{3K}$ ($D_{1K}\equiv r^{''}Y_{1K}$)
are defined in doubly-stretched coordinates
$r_i^{''}=(\omega_i/\omega_0)r_i$ ($i=x,y,z$)\cite{SK89}.
The coupling strengths of octupole interactions $\chi_{3K}$
are determined so as to reproduce the band-head
energies of the observed octupole vibrations
and standard values are used for the dipole strengths\cite{BM75}.
Where the band-head energies are unknown, we take
$\chi_{3K}=1.05(\chi_{3K})^{\rm HO}$, where $(\chi_{3K})^{\rm HO}$
indicate the self-consistent values obtained in the harmonic-oscillator
potential\cite{SK89}.

With RPA theory,
it is possible to investigate the correlations in heavier (deformed)
nuclei not contained in the mean field,
since excited states and their energies are obtained by
solving the RPA equation of motion, rather than by explicit
diagonalization.
After solving the RPA equation,
the total Hamiltonian (\ref{hamiltonian}) for even-even nuclei
at finite rotational frequency is written as
\begin{equation}
 \label{rpa-hamiltonian}
 H' = \hbox{const.} +
  \sum_{\alpha,n} \hbar\Omega_n^\alpha X_n^{\alpha\dagger}X_n^\alpha \ ,
\end{equation}
where $\alpha$ (=0,1) indicates the signature quantum numbers.
${X_n^\alpha}^\dagger$ and $\hbar\Omega_n^\alpha$ are the n-th
RPA-normal-mode creation operator and its excitation energy, respectively.

In addition to the energy eigenvalues,
the transition matrix elements from the one-phonon
to the vacuum state can be calculated within the RPA theory.
\begin{equation}
t[M(\lambda\mu)]\equiv \bra{\omega_{\rm rot}} M(\lambda\nu) \ket{n}
  = \bra{\omega_{\rm rot}}
    \left[ M(\lambda\nu), X_n^\dagger \right] \ket{\omega_{\rm rot}} \ ,
\label{transition-matrix}
\end{equation}
where $\ket{\omega_{\rm rot}}$ is the RPA vacuum at rotational
frequency $\omega_{\rm rot}$.
Since these quantities should be regarded as {\it intrinsic} values,
it is not trivial to calculate the transition amplitudes within this
model.
Marshalek developed a formula\cite{Mar75}
based on the lowest-order
boson expansion and ($1/J$)-expansion techniques on top of the cranking
model, which is valid in the high-spin limit.
The electric transition amplitudes are expressed in a simple form;
\begin{equation}
 \label{marshalek_formula}
 B(E\lambda; {\rm i}\rightarrow{\rm f})
 \approx
 \left| \bra{{\rm f}} M(\lambda\mu=\Delta J) \ket{{\rm i}} \right|^2 \ ,
\end{equation}
where $\Delta J = J_{\rm f} - J_{\rm i}$ and the transition operators
$M(\lambda\mu)$ are defined with respect to the rotation axis
($x$-axis).
The initial and the final state in
eq.(\ref{marshalek_formula}) are
$\ket{{\rm i}}=X_n^{\alpha\dagger}\ket{\omega_{\rm rot}}$ and
$\ket{{\rm f}}=\ket{\omega_{\rm rot}}$.

Shimizu and Nakatsukasa have recently shown\cite{SN95} that
the $J$-expansion theory combined with eq.(\ref{marshalek_formula})
usually leads to the same form as
the generalized intensity relations in Ref.\cite{BM75},
and enables us to microscopically calculate the intrinsic moments
entering in the relations.
Since these relations properly take into account the geometry of
the angular momentum vectors,
they are applicable to the low-spin states.
The intensity relation obtained for the electric ($E1$ and $E3$)
transitions is written as
\begin{eqnarray}
 \label{EL-GIR}
 \left[ B(E\lambda;{\rm i}\rightarrow {\rm f})^{\Delta J}
  \right]^{1/2}
 &\approx&  Q_t \,
 \langle J_{\rm i}, K_{\rm i},\ \lambda, -K_{\rm i}\ |
 \ J_{\rm f}, 0\rangle \,
 \left(1+q
 \left[J_{\rm f}(J_{\rm f}+1)-J_{\rm i}(J_{\rm i}+1)\right]\right)
  \hspace{1.5cm} \nonumber \\
 &+& Q_t' \, \sqrt{J_{\rm f}(J_{\rm f}+1)} \,
 \langle J_{\rm i}, K_{\rm i},\ \lambda, -K_{\rm i}-1\
 |\ J_{\rm f}, -1\rangle \ ,
\end{eqnarray}
where we make use of the fact
that $K$-quantum number of the ground state band equals zero
($K_{\rm f}=0$).
Note that the parameter $Q_t'$ has non-zero value only for $E3$
transitions of $K=1$ and $K=2$ octupole vibrations.
{}From the calculated transition amplitudes (\ref{transition-matrix}) and
their derivative with respect to $\omega_{\rm rot}$,
the parameters ($Q_t, q, Q_t'$) in eq.(\ref{EL-GIR})
can be obtained\cite{SN95}.


\section{I\lowercase{nterplay between rapid rotation and octupole
 vibrations}}
\label{sec: three}

In this section, we present general properties of
octupole vibrations at high angular momentum.
We classify the responses of the phonon to rotation as
``phonon alignment'' (to the rotation axis), ``phonon breakdown'',
and ``phonon disappearance''.

The ground states of even-even deformed nuclei are normally in
the superconducting phase and there is an energy gap,
$\Delta$, which is typically about 1 MeV for
both neutrons and protons.
Therefore the energy of the lowest excitation mode
might be expected to be about $2\Delta\approx 2$ MeV.
However, the attractive ($T=0$) residual interactions significantly lower
the energies of some excited states,
and give rise to the low-lying vibrational states.

At low spin, where residual interactions are more important
than the Coriolis force,
octupole vibrations are well characterized by
their $K$-quantum numbers
(the projection of angular momentum along the symmetry axis).
At higher spin, the Coriolis field mixes
states with different $K$
and leads to make the angular momentum of the octupole phonon
align along the rotational axis ({\it phonon alignment}).
Since octupole phonons carry 3 units of angular momentum,
one can expect the maximum alignment will be $i_{\rm max}\approx 3\hbar$.
Coriolis mixing is favored by large rotational energy,
and unfavored by the quadrupole-deformation splitting of the
single-particle energies.
Thus, generally speaking, phonon alignment is expected to occur more
easily in nuclei with smaller $\beta$ (deformation).

\begin{figure}[thb]
\epsfysize=0.4\textheight
\centerline{\epsfbox{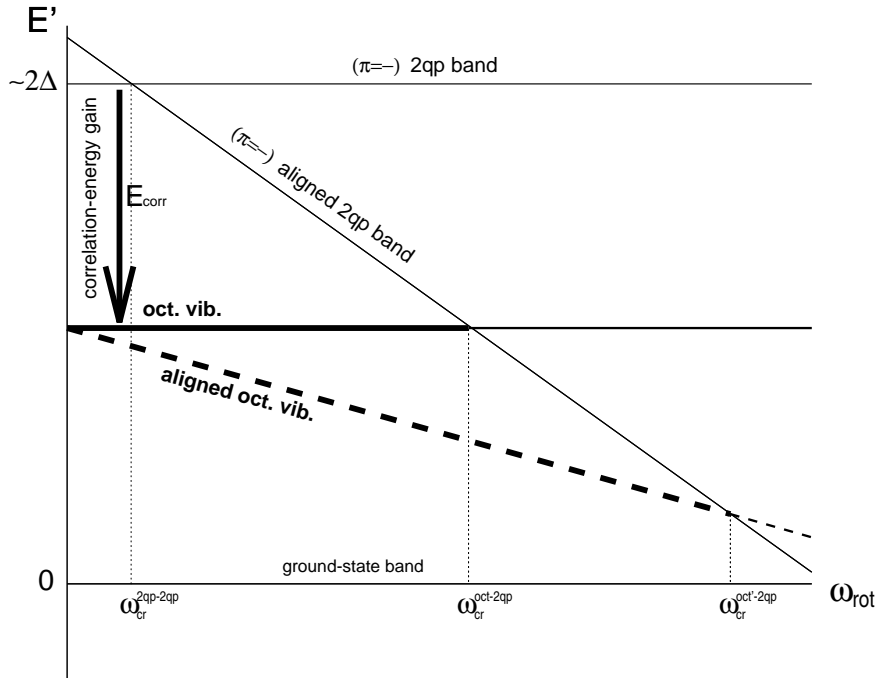}}
\caption{
Schematic routhians for octupole vibrations and negative-parity
two-quasiparticle states.
See text for explanations.
}
\label{one}
\end{figure}

This phonon-alignment phenomenon has only been observed in a few
deformed nuclei because the octupole phonons are usually
broken-down at high spin.
As is explained in the previous section,
the phonon excitations consist of linear combinations of many
two-quasiparticle states (in the RPA order).
Some individual quasiparticles associated with high-$j$ intruder orbits
feel the strong Coriolis force and try to align their angular momenta
parallel to the rotational axis ($i_{\rm 2qp}\approx 5\sim 10\hbar$).
At low spin, these individual alignments are prevented by the
octupole correlations, which try to couple two-quasiparticles to spin
$3\hbar$.
If, however, at high spin the Coriolis field becomes strong enough,
then, the individual aligned two-quasiparticles
are released from the octupole coupling ({\it phonon breakdown}).
\begin{equation}
\ket{{\rm oct.vib.}}=\sum_{k} C_k \ket{k}_{\rm 2qp}\quad
\longrightarrow\quad \ket{n}_{\rm aligned-2qp} \ .
\end{equation}
This can be regarded as a crossing between the octupole vibration and
the aligned two-quasiparticle state (see Fig.\ref{one}).
One can make a rough estimate for the critical (crossing) frequency,
$\hbar\omega_{\rm cr}^{\rm oct-2qp}
= E_{\rm corr}/(i_{\rm 2qp} - i_{\rm oct})$.
As is seen in Fig.\ref{one}, if the octupole vibration is rotationally
aligned ($i_{\rm oct}\approx 3\hbar$),
$\omega_{\rm cr}^{\rm oct-2qp}$
becomes much larger than in the non-aligned case ($i_{\rm oct}\approx 0$).

When pairing correlations are weakened at higher spin,
octupole vibrations will generally become less collective.
Furthermore, in normal-deformed nuclei,
most low-energy negative-parity two-quasiparticle states are built by
occupying one normal-parity level and
one high-$j$ intruder level (e.g., $j_{15/2}, i_{13/2}$).
Thus, octupole vibrations lose their collectivity after the crossings
(release of high-$j$ orbits),
and eventually vanish ({\it phonon disappearance}).

However, in SD nuclei, those two-quasiparticles are not necessarily
associated with high-$j$ orbits,
because each major shell consists of almost equal numbers of positive-
and negative-parity levels which are degenerate at the 2:1 axis ratio
(without the spin-orbit potential).
Therefore one can expect that octupole vibrations built on a
SD shape will be more stable against the crossings of aligned
two-quasiparticle states.
This provides us with a better chance to observe the high-spin
octupole vibrations in SD nuclei.

\section{O\lowercase{ctupole vibrations in} $^{164}$Y\lowercase{b}}

In this section the calculated results for the octupole vibrations
in the rare-earth nucleus ($^{164}$Yb) are presented;
in this nucleus we can see the various
rotational effects discussed in the previous section.
The lowest negative-parity bands with signature $\alpha=0$ and 1
are assigned as the $K^\pi=4^-$ and $5^-$ bands in Ref.\cite{NDS92}.
However the alignment plots shows a sharp rise at $J\approx 8\hbar$
for the $\alpha=0$ band which we interpret
as a crossing between an octupole
band and an aligned two-quasiparticle band\cite{Jon86}.

\begin{figure}[tbhp]
\epsfysize=0.4\textheight
\centerline{\epsfbox{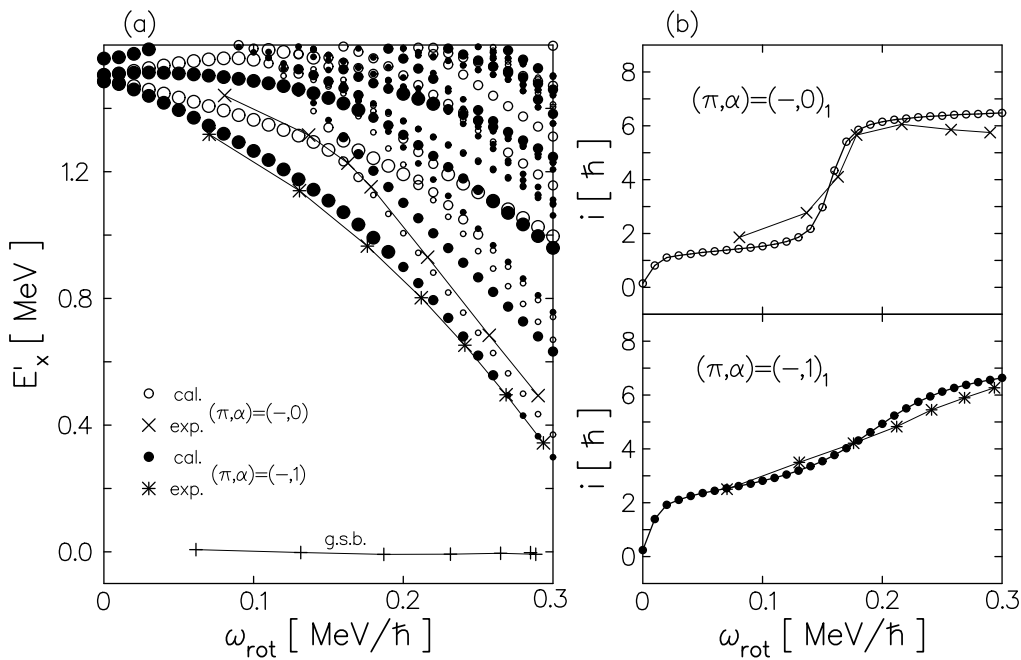}}
\caption{
Results of RPA calculations for $^{164}$Yb.
Quadrupole deformation $\epsilon=0.23$, pairing gaps
$\Delta_n=1.0$MeV, $\Delta_p=0.8$MeV
(at $\omega_{\rm rot}=0$) were used in the calculation.
(a) Calculated negative-parity RPA eigenvalues plotted as a function of
rotational frequency.
Large, medium, and small circles indicate
RPA solutions with E3 transition amplitudes
$\left[ \sum_K\left| \bra{n} Q_{3K}^e \ket{0}\right|^2 \right]^{1/2}$
larger than 200 efm$^3$,
larger than 100 efm$^3$ and less than 100 efm$^3$, respectively.
(b) Plot of aligned angular momentum as a function of rotational
frequency
for the lowest negative-parity state with
$\alpha=0$ (upper) and $\alpha=1$ (lower).
}
\label{two}
\end{figure}

In Fig.\ref{two}(a), the calculated RPA eigenvalues are shown with circles
whose size qualitatively indicates their $E3$ amplitudes.
The calculated $K=1$, 2 and 0 band-head states are almost degenerate in
energy ($E_x\sim 1.5$ MeV).
The lowest $\alpha=1$ octupole state is rotationally aligned at low spin,
and then gradually loses its vibrational character around
$\omega_{\rm rot}\approx 0.2\mbox{MeV}/\hbar$
becoming the aligned two-quasineutron state
($\nu (i_{13/2}\otimes h_{9/2})^{\alpha=1}$).
For this case, it is difficult to identify a specific crossing partner
because the octupole strength is shared by several states
after the crossing.
The lowest $\alpha=0$ octupole state is
crossed by a two-quasiparticle state at
$\omega_{\rm rot}\approx 0.16\mbox{MeV}/\hbar$, which corresponds to
the crossing illustrated in Fig.\ref{one}.
Most octupole vibrations are calculated to lose their
collectivity after the crossings
and vanish at $\omega_{\rm rot}=0.3\mbox{MeV}/\hbar$.

The corresponding alignment plot is shown in Fig.\ref{two}(b).
The experimental characteristics are well accounted for by the calculation.
It is worth mentioning that the sharp alignment gain and the large
signature splitting can not be explained in calculations which do not
include octupole correlations.

\section{O\lowercase{ctupole vibrations in} $^{238}$U}

In $^{164}$Yb the lowest octupole state is rotationally aligned
according to our calculation,
however, this alignment is not so clear in the experiment
because there are only $2\sim 3$ data points before the crossing.
This is because the octupole-correlation energy
and moment of inertia are both small\cite{Vog70}.
In this context actinide nuclei should be better
suited for the observation of octupole-phonon alignment.

\begin{figure}[tbhp]
\epsfysize=0.38\textheight
\centerline{\epsfbox{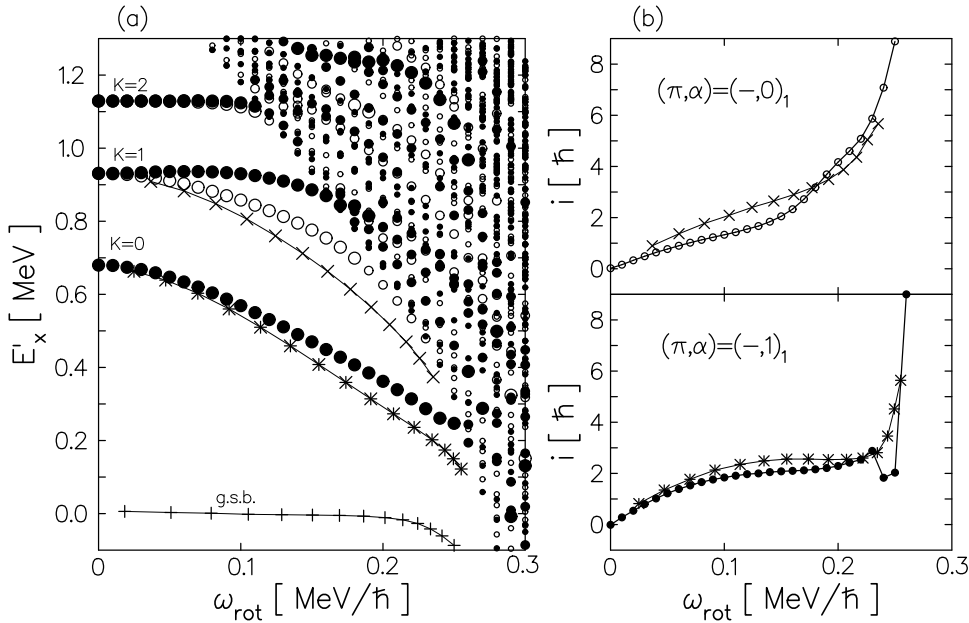}}
\caption{
As Fig.\protect\ref{two}, but for $^{238}$U.
$\epsilon=0.23$,
$\Delta_n=0.7$MeV, $\Delta_p=0.8$MeV are assumed to be constant
with rotational frequency.
Large, medium, and small circles indicate
$\left[ \sum_K\left| \bra{n} Q_{3K}^e \ket{0}\right|^2 \right]^{1/2}$
larger than 300 efm$^3$,
larger than 150 efm$^3$ and less than 150 efm$^3$, respectively.
}
\label{three}
\end{figure}

The lowest octupole band ($\alpha=1$) in $^{238}$U
has been known up to $J^\pi=19^-$ for a long time\cite{Gro75}
and corresponding theoretical calculation has been done\cite{Vog70,RER86}.
Recently experiments at Chalk River by Ward et al.\cite{War95}
have extended it up to $J^\pi=31^-$.
These same experiments also observed the lowest
$\alpha=0$ octupole band to $J^\pi=28^-$.
These experimental data are compared with the calculated results
in Fig.\ref{three}.

The rotationally-aligned-octupole phonon is clearly seen for
the lowest $\alpha=1$ octupole band
at $\omega_{\rm rot}\geq 0.1\mbox{MeV}/\hbar$ in which
the observed aligned angular momentum is about $2.5\hbar$.
This octupole state is suddenly broken-down at
$\omega_{\rm rot}\approx 0.25\mbox{MeV}/\hbar$ and changed into
an aligned two-quasineutron state.
This phenomenon is seen in Fig.\ref{three} (b) as a sharp alignment gain.
The $\alpha=0$ octupole phonon has similar properties,
although the process of losing the octupole vibrational character is
more gradual.

The characters of the alignment process
({\it gradual} or {\it sudden})
between the $\alpha=0$ and 1 octupole states in $^{238}$U
is opposite to that in $^{164}$Yb.
In addition, the octupole vibrations quickly disappear
after the crossings.
We attribute this to the high density of aligned two-quasiparticle
states in $^{238}$U compared with $^{164}$Yb.

Since branching ratios for the lowest $\alpha=1$ band
have been measured up to $J^\pi=19^-$ in the new experiment,
validity of formula (\ref{EL-GIR}) can be tested in the regions of
low and medium spin.
In Fig.\ref{four}, $E1$ branching ratios
\begin{equation}
\label{D-over-D}
\left[\frac{B(E1;[0^-, J]\rightarrow [0^+, J+1])}
{B(E1;[0^-, J]\rightarrow [0^+, J-1])}\right]^{1/2}
\left| \frac{\langle J,\ 0,\ 1,\ 0\ |\ J-1,\  0\rangle}
{\langle J,\ 0,\ 1,\ 0\ |\ J+1,\ 0\rangle}\right|
= \frac{1+2(J+1)q}{1-2Jq}\ ,
\end{equation}
\begin{minipage}[t]{0.4\textwidth}
\begin{figure}[tbhp]
\vspace*{-0.7cm}
\epsfxsize=0.9\textwidth
\centerline{\epsfbox{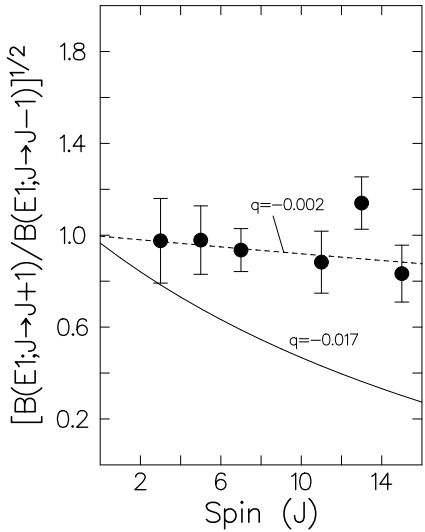}}
\caption{
E1 branching ratios, l.h.s. of eq.(\protect\ref{D-over-D}),
for the lowest octupole state in $^{238}$U.
Dotted lines indicate r.h.s. of eq.(\protect\ref{D-over-D})
with $q=-0.002$.
Solid lines are predicted by calculations ($q=-0.017$).
Experimental data (circles) are taken from Ref.\protect\cite{War95}.
}
\label{four}
\end{figure}
\end{minipage}
\hspace{0.5cm}
\begin{minipage}[t]{0.55\textwidth}
are shown.
The results indicate that
Coriolis mixing effects on
these $B(E1)$ branching ratios are very small ($q\approx 0$).
This is because the $E1$ strengths of the $K=0$ octupole band are
about 10 times larger than the $K=1$ band.
Thus, even if the $K=0$ octupole states
were strongly contaminated by $K=1$ states,
it would have little affect on the $B(E1)$ data.
The calculation underestimates the relative difference of $E1$ strength
between $K=0$ and 1, which results in an overestimation
of the $q$ parameter.

\hspace*{0.6cm}By contrast,
$E3$ strengths are significantly affected by the Coriolis coupling
because the $K=0$ and 1 states have similar $E3$ strengths.
Although the $E3$ data at high spin are not available,
the effect can be seen in the spin $3^-$ octupole states.
Table \ref{E3} shows the results obtained by using eq.(\ref{EL-GIR})
with and without the Coriolis-coupling term ($q,Q_t'$).
The concentration of $E3$ strengths onto the lowest state\cite{NV70}
is obtained.
\end{minipage}

\mediumtext
\begin{table}[tbhp]
\begin{center}
\begin{minipage}{0.6\textwidth}
\begin{tabular}{c|cc|cc}
\hline
 & \multicolumn{2}{c}{Theory}  & \multicolumn{2}{c}{Experiment}\\ \hline
$K$, $J^\pi$ & $B(E3; 3^- \rightarrow 0^+)$ & $B(E3)$
                                            & $B(E3)$ & $E_x(3^-)$ \\
             & LIC                      & GIC  & & \\
    & [W.u.] & [W.u.] & [W.u.] & [keV]\\ \hline
0, $3^-$ & 18.5 & 25.2 & 24.2  & 732\\
1, $3^-$ & 20.1 & 17.5 & 7.8   & 998\\
2, $3^-$ & 12.1 & 9.6  & 7.1   & 1169 \\
3, $3^-$ & 9.7  & 8.3  &       & \\ \hline
\end{tabular}
\end{minipage}
\caption{Comparison of $B(E3; 3_{\rm oct.}^- \rightarrow 0_g^+)$ for
$^{238}$U calculated with the RPA
in the leading-order intensity calculation (LIC) and
in the generalized intensity calculation (GIC).
The Coriolis coupling is taken into account in the lowest order in GIC
while it is neglected in LIC.
Experimental values are taken from Ref.\protect\cite{MM94}}
\label{E3}
\end{center}
\end{table}

\widetext
\section{O\lowercase{ctupole vibrations in}
SD $^{190,192,194}$H\lowercase{g}}

The importance of octupole correlations in SD states have been
suggested by many authors\cite{SDoct}.
As discussed in section \ref{sec: three},
octupole vibrations built on the SD yrast band
might be more stable at higher angular momentum than those built on
normal states.
Since pairing correlations enhance the
collectivity of octupole vibrations,
the systematic observation of collective octupole vibrations
may be more likely in the A=190 region compared to the A=150 region.

\begin{figure}[tbhp]
\vspace*{-0.7cm}
\epsfxsize=0.9\textwidth
\centerline{\epsfbox{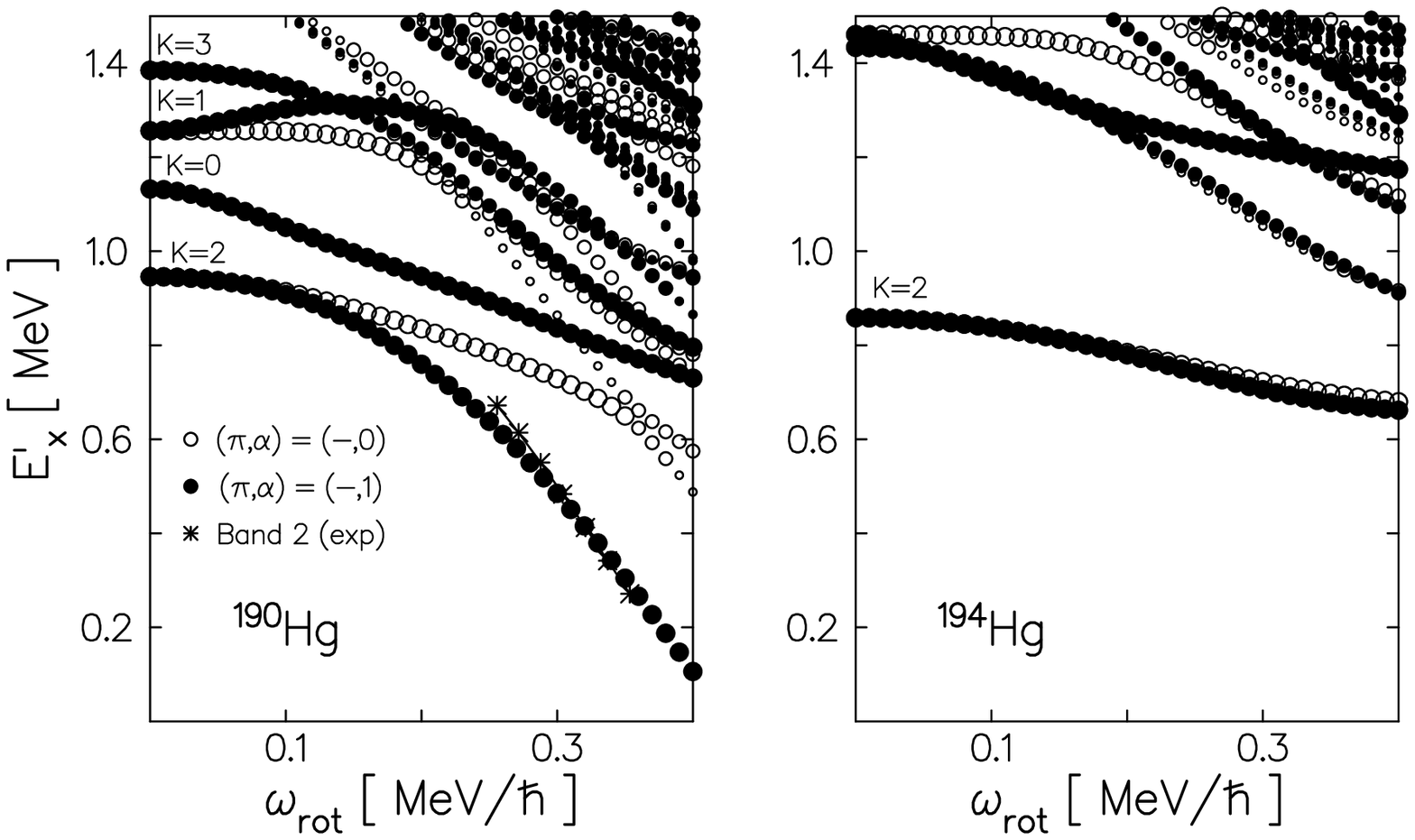}}
\caption{
Calculated negative-parity RPA eigenvalues
for SD $^{190}$Hg (left) and $^{194}$Hg (right).
$\epsilon=0.44$, pairing gaps
$\Delta_n=0.8$MeV, $\Delta_p=0.6$MeV
are assumed to be constant.
Routhians for the yrast SD band correspond to the horizontal axis
($E'_x=0$).
Experimental routhians ($\ast$) for Band 2 in $^{190}$Hg are
taken from Ref.\protect\cite{Cro94}.
See caption to Fig.\protect\ref{two}(a).
}
\label{five}
\epsfxsize=0.9\textwidth
\centerline{\epsfbox{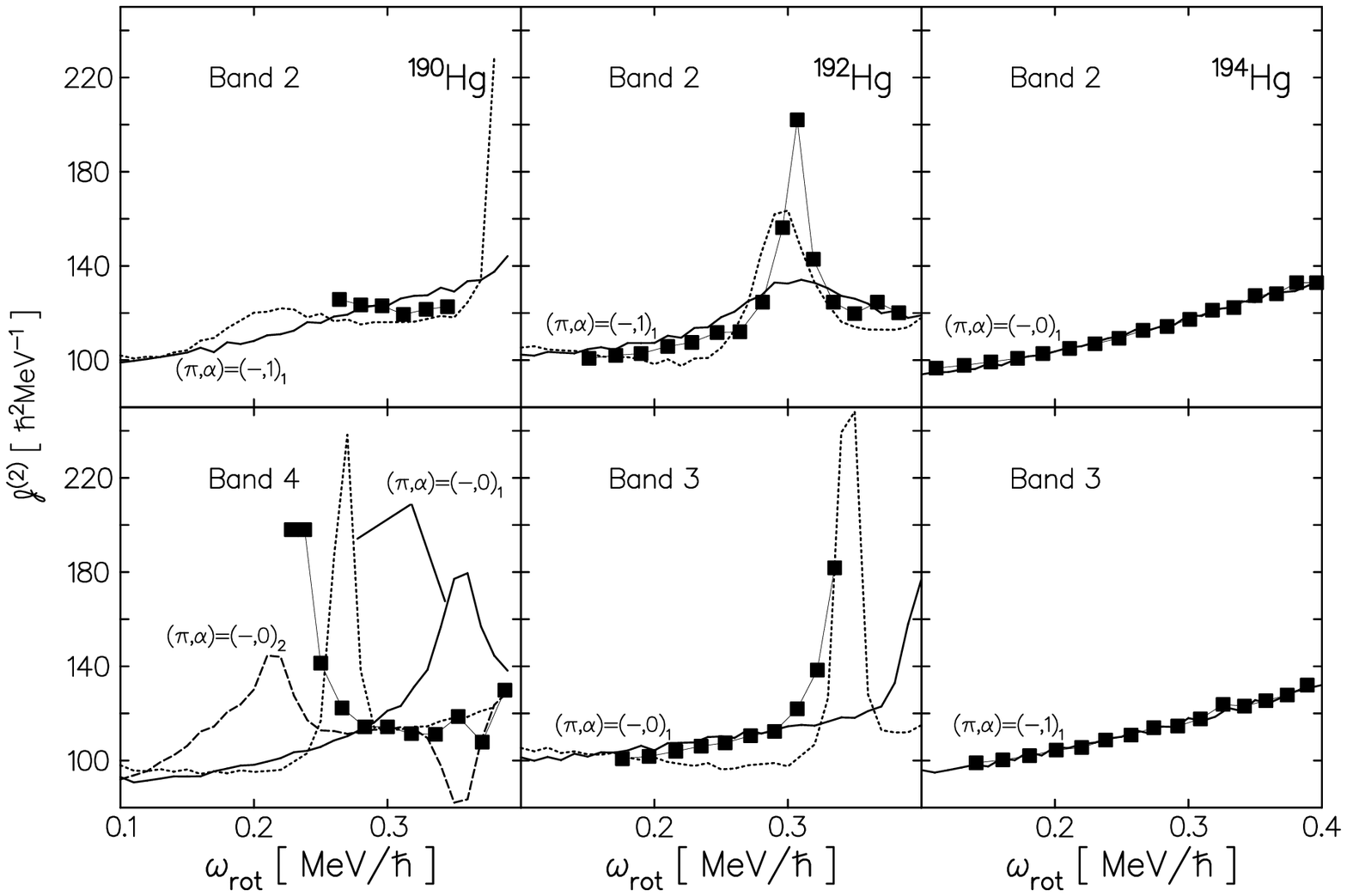}}
\caption{
Calculated (lines) and experimental (symbols) dynamical moments of
inertia
for $^{190}$Hg (left), $^{192}$Hg (middle) and $^{194}$Hg (right).
Experimental data are taken from Ref.\protect\cite{Cro94,SDexp}.
See text for detail.
}
\label{six}
\end{figure}

The calculation predicts that $K=2$ octupole vibrations are
the lowest for $^{190,192,194}$Hg ($E_x\approx 1$MeV).
In Fig.\ref{five}, calculated RPA routhians are
shown for $^{190}$Hg and $^{194}$Hg.
In the calculations the routhians behave differently
at higher frequency:
In $^{194}$Hg, both signatures of the lowest octupole routhians
are almost constant with rotational frequency.
In $^{190}$Hg, the lowest $\alpha=1$ octupole vibration is rotationally
aligned at $\omega_{\rm rot} > 0.2\mbox{MeV}/\hbar$,
while the $\alpha=0$ state is not and is crossed by the aligned
two-quasineutron state at $\omega_{\rm rot}\approx 0.35\mbox{MeV}/\hbar$.
This is a good example of how an aligned octupole phonon
($\alpha=1$) can survive up to higher frequency than can a non-aligned
phonon ($\alpha=0$), as is seen in Fig.\ref{one}.
In $^{192}$Hg, both signatures are unaligned,
and are crossed by two-quasineutron bands at
$\omega_{\rm rot}=0.3\sim 0.4\mbox{MeV}/\hbar$.
This is a similar behavior to the $\alpha=0$ band in $^{190}$Hg.

Band 2 in $^{190}$Hg has been interpreted as an octupole vibrational band
in Ref.\cite{Cro94}.
We will now show that octupole bands may be more prevalent than supposed.
Calculated dynamical moments of inertia for $^{190,192,194}$Hg
are compared with available experimental data in Fig.\ref{six}.
Solid lines are results calculated with constant pairing gaps while
dotted lines are calculated with pairing gaps dynamically
reduced with frequency
following a phenomenological prescription\cite{Wys90}.
For $^{192}$Hg, the sharpness and positions of the peaks in
${\cal J}^{(2)}$
are much improved by the reduced pairing.
A more detailed discussion of calculations with the dynamically reduced
pairing will be given in another paper\cite{Nak96}.
For Band 4 in $^{190}$Hg (lower left panel in Fig.\ref{six}),
there are two possible candidates.
One is the lowest $\alpha=0$ state and the other is the second lowest
$\alpha=0$ state.
The second one (dashed lines)
shows much better agreement
if constant pairing is used.
However the lowest one becomes better with
reduced pairing.

Peaks observed for Band 4 in $^{190}$Hg,  Bands 2 and 3 in $^{192}$Hg
can not be reproduced by the mean-field calculations
(quasiparticle routhians in the Nilsson or Woods-Saxon potential).
The crossing frequency $\omega_{\rm cr}^{\rm 2qp-2qp}$ between aligned and
non-aligned two-quasiparticle states is predicted to be around
$0.1\mbox{MeV}/\hbar$, which is much earlier than frequency of the
observed peaks.
Since octupole vibrations gain the correlation energy,
their total energy can be much lower than $2\Delta$,
and the crossing frequency $\omega_{\rm cr}^{\rm oct-2qp}$ is
significantly delayed
(see Fig.\ref{one}), giving a good agreement with experiment.

We have shown in the RPA calculation that the $E1$ decays of octupole
bands are very sensitive to nuclear structure,
so that decays of an octupole band will not necessarily be observed
as in the $^{190}$Hg case.
The results obtained by using formula (\ref{marshalek_formula})
suggest $B(E1; J\rightarrow J-1)$ equals $10^{-5}$
to $10^{-4}$W.u.
at $\omega_{\rm rot}=0.25\mbox{MeV}/\hbar$ for Band 2 in $^{190}$Hg,
and for the other bands $10^{-9}$ to $10^{-6}$W.u.
Although the calculated $B(E1)$ value for Band 2 in $^{190}$Hg
is smaller than the experimental value ($\sim 10^{-3}$W.u.) \cite{Cro94},
it is shown to be much larger than
the $B(E1)$ value of the other bands.

\section{S\lowercase{ummary}}

Octupole vibrations at high angular momenta are investigated by
microscopic calculations based on the cranked shell model extended
by RPA.
Various properties produced by rapid rotation are discussed,
e.g., octupole-phonon alignment, phonon breakdown, and phonon disappearance.
For excited SD bands in $^{190,192,194}$Hg, a new interpretation
based on the octupole vibrational bands is proposed,
which shows a good agreement with observed features.

\medskip

I gratefully acknowledge the contribution of many colleagues
from both theoretical and experimental sides.
Especially, I thank K.~Matsuyanagi, S.~Mizutori,
W.~Nazarewicz, Y.R.~Shimizu and D.~Ward for collaborations
and valuable discussions.


\end{document}